\documentclass[aps,prb]{revtex4}
\usepackage{amsmath,amssymb}
\usepackage{graphicx}
\usepackage{dcolumn}
\usepackage{bm}
\usepackage{color}
\usepackage{mathtools}

\newcommand{\mbb}{\mathbb}
\newcommand{\mc}{\mathcal}

\newcommand{\tet}{\texttt}
\newcommand{\pr}{\partial}

\newcommand{\sign}{\text{sign}}

\begin{document}

\title{Finite-temperature plasmons, damping and collective behavior for $\alpha - \mc{T}_3$ model}

\author{
Andrii Iurov$^{1}\footnote{E-mail contact: aiurov@mec.cuny.edu, theorist.physics@gmail.com}$,
Liubov Zhemchuzhna$^{1,2}$,
Godfrey Gumbs$^{2,3}$, 
Danhong Huang$^{4}$,
Dipendra Dahal$^{5}$,
Yonatan Abranyos$^{2}$
}

\affiliation{
$^{1}$Department of Physics and Computer Science, Medgar Evers College of the City University of New York, Brooklyn, NY 11225, USA\\
$^{2}$Department of Physics and Astronomy, Hunter College of the City University of New York, 695 Park Avenue, New York, NY 10065, USA\\
$^{3}$Donostia International Physics Center (DIPC), P de Manuel Lardizabal, 4, 20018 San Sebastian, Basque Country,
Spain\\
$^{4}$Air Force Research Laboratory, Space Vehicles Directorate, Kirtland Air Force Base, NM 87117, USA\\
$^{5}$Department of Physics, University of Houston, 4800 Calhoun Rd, Houston, TX 77004 USA
}

\date{\today}

\begin{abstract}
We have conducted a thorough theoretical and numerical investigation of the electronic susceptibility, polarizability, plasmons, their damping rates, as well as the static screening in pseudospin-1 Dirac cone materials with a flat band, or for a general $\alpha - \mc{T}_3$ model, at finite temperatures. This includes calculating the polarization function, plasmon dispersions and their damping rates at arbitrary temperatures and obtaining analytical approximations the long wavelength limit, low and high temperatures. We demonstrate that the integral transformation of the polarization function cannot be used directly for a dice lattice revealing some fundamental properties and important applicability limits of the flat band dispersions model. At $k_B T \ll E_F$, the largest  temperature-induced change of the polarization function and plasmons comes from the mismatch between the chemical potential and the Fermi energy. We have also obtained a series of closed-form semi-analytical expressions for the static limit of the polarization function of an arbitrary $\alpha - \mc{T}_3$ material at any temperature with exact analytical formulas for the high, low and zero temperature limits which is of tremendous importance for all types of transport and screening calculations for the flat band Dirac materials.  
\end{abstract}

\maketitle

\section{Introduction}
\label{s01}

Plasmons, the self-sustained  quantum oscillations of  electrons in a conducting material\,\cite{pines2018elementary, nozieres1958electron}, have demonstrated considerable potential for various technological applications and have become an important  research subject for any newly discovered low-dimensional structure, such as graphene.\,\cite{luo2013plasmons,jablan2009plasmonics,jablan2013plasmons} The nature of the plasmon dispersion is directly related to the low-energy electron dispersion in a given material. For example, all the two-dimensional materials including the two-dimensional electron gas (2DEG) demonstrate $\backsim \sqrt{q}$ plasmon dispersion  in the long wavelength limit, but their dependence on the electron density may reveal the fingerprints of Dirac cone material ($\backsim \sqrt{n}$) and even the presence of a bandgap or an additional dispersionless band in the single-electron energy spectrum.\,\cite{wunsch2006dynamical, sarma2013intrinsic} When a plasmon frequency reaches THz or visible light range, a polariton quasiparticle could be formed due to plasmon-photon strong coupling.\,\cite{berini2012surface,brar2014hybrid,iurov2017effects}

\medskip 
\par
More generally, plasmonics or the study of the generation and propagation of the plasmon excitations in conducting media, surfaces and the interfaces between a metal and a dielectric material, has become one of the most promising research directions in both fundamental  and applied material science.\,\cite{grigorenko2012graphene, gonccalves2016introduction, garcia2014graphene, ju2011graphene,ni2018fundamental} It has also been gaining much attention by experimentalists waning to improve fabrication techniques of these nanometer-size materials and their  utilization as plasmonic devices.\,\cite{yan2012tunable, politano2013unravelling} The primary areas of plasmonics include nanophotonics\,\cite{catchpole2011plasmonics, monticone2017metamaterial,wei2012nanowire}, quantum  plasmonics\,\cite{tame2013quantum, lundeberg2017tuning, jacob2011plasmonics} plasmonic integrated circuits and waveguides\,\cite{sorger2012toward, salvador2008analysis, kim2011graphene,bozhevolnyi2008plasmonic}, and optical manipulation of new types of plasmon generators\,\cite{yao2014efficient, constant2016all,yao2018broadband,de2017plasmon}.

\medskip
\par

Evidently,  a clear understanding of the plasmon behavior, including its energy range and dispersion relation, is the key information needed for successful design  of such mentioned  plasmonic devices. The investigation of a quantum plasmon oscillation is based on the dielectric function which is closely related to  the so-called polarization function\,\cite{mahan2010condensed, fetter2012quantum} that is the central object of our study in this paper. The polarization function is related to the crucial phenomena and properties of a low-dimensional material, such as quantum transport, Boltzmann conductivity, static screening of a charged or magnetic impurity,\,\cite{sobol2016screening, galperin2018entropy, gusynin2006magneto} phonon excitation spectrum and collective effects.\,\cite{hwang2009screening, sarma2011electronic} Therefore, the polarization function was obtained and analyzed in all innovative two-dimensional  materials\,\cite{hofmann2016surface}: graphene\,\cite{hwang2007dielectric, wunsch2006dynamical, hwang2008screening,  pyatkovskiy2008dynamical,pyatkovskiy2011dynamical,gamayun2011dynamical}, silicene\,\cite{tabert2014dynamical,van2014spin,van2021plasmon, wu2014temperature}, molybdenum disulfide\,\cite{scholz2013plasmons} and other lattices with spin-orbit coupling\,\cite{scholz2012dielectric}, anisotropic Dirac materials\,\cite{xu2021optical, ahn2021theory, hayn2021plasmons} and, most recently, in $\alpha - \mc{T}_3$ materials and a dice  lattice\,\cite{Malc01, huang2019interplay}. Specifically,  careful attention has been give to the zero-frequency static limit of the polarization function which is critical for electron scattering and transport  calculations\,\cite{Malc01,hwang2009screening}

\medskip 
\par
As mentioned above, the electron doping density $n$ also plays a crucial role in shaping out its plasmon dispersion and other collective  properties of a specific material. In this regard, we separate doped (extrinsic) materials with a finite chemical potential from intrinsic ones with the Fermi energy located exactly at the Dirac point. At zero temperature, only doped Dirac materials can support a plasmon but  at finite temperature, the thermal population of electrons and holes plays the role of effective doping and such intrinsic plasmon become possible\,\cite{sarma2013intrinsic} At high temperatures, even large doping does not play a crucial role since the effect of the temperature is essentially similar to that  of  finite doping so that intrinsic and extrinsic materials behave essentially in the same way. The temperature dependence of the chemical potential is also an important characteristic of a lattice which can also reveal its bandstructure\,\cite{hwang2009screening,gumbs2017combined,iurov2017thermal}

\medskip 
\par
A family of $\alpha - \mc{T}_3$ lattices seems the most promising type of novel Dirac cone materials\,\cite{zhou2018dynamical,tan2021anisotropic} and a potential replacement for graphene due to their revolutionary new low-energy electronic bandstructure. Their atomic build up is such that there is an extra atom in their  honeycomb lattice. As a result, there exist additional electronic hopping coefficients so that their electronic spectrum acquires an  extra flat or dispersionless band exactly at the Dirac point between the valence and conduction bands.   Therefore,  the $\alpha - \mc{T}_3$ model and its electronic properties could be described by pseudospin-1 Dirac-Weyl Hamiltonian with a phase  angle $\phi = \tan^{-1} \alpha$ , where $\alpha$ denotes the relative strength of the additional electron hopping coefficient associated with  the hub atom.  The case with $\alpha = 0$ corresponds to graphene with a set of decoupled hub atoms in the center of each hexagon, while $\alpha = 1$  represent a dice lattice with a maximum effect of these central atoms. Such a unique modification of the Dirac cone energy dispersions results in surprisingly new and unusual electronic\,\cite{illes2017properties, gorbar2019electron, iurov2020klein, wang2021super,  xu2021klein, weekes2021generalized, dey2019floquet, dey2018photoinduced}, optical \,\cite{bryenton2018optical, iurov2020quantum, chen2019enhanced, iurov2019peculiar}, magnetic \,\cite{islam2017valley, raoux2014dia, illes2016magnetic, oriekhov2021current, islam2018driven,biswas2016magnetotransport} and collective \,\cite{biswas2018dynamics, iurov2020many, Malc01} properties of such lattices which have become one of the most important subjects in  modern condensed matter physics. At this point there is strong evidence of possibly fabricating various $\alpha - \mc{T}_3$  materials and dice lattices with a flat band\,\cite{leykam2018artificial, wang2011nearly}.

\medskip 
\par 

The remainder of this paper is organized as follows: in Section \ref{s02}, we review the theoretical formalism for calculating the zero- and finite-temperature polarization functions in flat-band Dirac materials  with an emphasis on the approximations corresponding to the high- and low-temperature limits. Section \ref{s03} is concerned with finding the long wavelength limit of the polarization function for both zero and finite temperatures. Finally, we perform a thorough calculation of the static limit of the polarization function, which is also referred to as Lindhard function, for all $\alpha - \mc{T}_3$ materials at any temperature in Section \ref{s04}, including zero- and the low-temperature limits. The detailed derivations of the general formula, the long wavelength limit and the static limit of the temperature-dependent polarization functions are provided in Appendices \ref{AppA}, \ref{AppB} and \ref{AppC}.

\section{Polarization function in $\alpha - \mc{T}_3$ materials at  finite temperature: general formalism}
\label{s02}

\par 
The low-energy Hamiltonian in $\alpha-\mc{T}_3$ is written in the following form

\begin{equation}
\label{H0}
\hat{\mc{H}}_{\alpha} (\mbox{\boldmath$k$} \, \vert \, \tau, \phi) = \hbar v_F \,
\left[
\begin{array}{ccc}
0 & k^\tau_- \cos \phi &  0 \\
0 & 0 & k^\tau_-\sin \phi \\
0 & 0 & 0
\end{array}
\right]
+ h.c.\ ,  
\end{equation}
where the  geometric phase $\phi=\tan^{-1}\alpha$ and $k^\tau_\pm=k^\tau_x\pm i k^\tau_y$ are obtained from the components of the electron wave vector ${\bf k}$.

\medskip 

One of the most important features in connection with plasmons at zero and finite temperatures is 
its dispersion relations, i.e., dependence of the plasmon frequency $\omega$ on wave number $q$. Physically, these complex relations can be determined from the zeros of a dielectric function $\epsilon_T(q,\,\omega)$ given by

\begin{equation}
\epsilon_{\phi,T} (q,\,\omega) = 1 - v(q) \, \Pi_{T}(q, \omega \, \vert \, \mu(T)) = 0 \ ,
\label{eps01}
\end{equation}
where $v(q)=2 \pi \alpha_r/q \equiv e^2 /2\epsilon_0\epsilon_r q$ is the 2D Fourier-transformed Coulomb potential, 
$\alpha_r = e^2/4\pi\epsilon_0 \epsilon_r$, and $\epsilon_r$ represents the dielectric constant of the host material.

\medskip  
The dielectric function introduced in Eq.\,\eqref{eps01} is determined directly by the finite-temperature {\it polarization function},
or {\it polarizability}, $\Pi_{T}(q,\omega \, \vert \, \mu(T))$, which is, in turn, related to its {\it zero-temperature counterpart}, $\Pi_{0} (q,\omega \, \vert \, E_F)$, by an integral convolution with respect to different Fermi energies\,\cite{maldague1978many}, given by

\begin{equation}
\Pi_{T}(q, \omega \, \vert \, \mu(T)) = \frac{1}{k_B T} \, \int\limits_{0}^{\infty}  d\eta \,
\frac{\Pi_{0} (q,\omega \, \vert \, \eta)}{1 + \cosh \left[(\mu - \eta) / k_B T \right] } \ ,
\label{Pi0T}
\end{equation}
where the integration variable $\eta$ stands for the electron Fermi energy at $T=0$. This equation is derived for electron doping with $\eta=E_F>0$. We note that, in order to evaluate this integral, one needs to know in advance  how the chemical potential $\mu(T)$ varies with temperature $T$. 

\medskip

In RPA, the polarization function for a pseudospin-1 Hamiltonian of $\alpha-\mc{T}_3$ model is given by 

\begin{equation}
\Pi_0(q, \omega \, \vert \, \phi, \mu(T)) = \frac{g}{4 \pi^2} \, \int d^2{\bf k} \sum\limits_{\gamma,\gamma' = 0 \, \pm 1} \, 
\mbb{O}_{\gamma_1,\gamma_2}(\beta_{{\bf k},{\bf k}+{\bf q}}) \, \frac{n_F[\epsilon_\gamma (k), \mu] - n_F[\epsilon_{\gamma'} (\vert {\bf k} +
\bf{q} \vert), \mu]}{ \left(\hbar \omega + i 0^+ \right) + \epsilon_\gamma (k) - \epsilon_{\gamma'} (\vert {\bf k} + \bf{q} \vert)} \, ,
\label{Pi00}
\end{equation}
where $n_F[\epsilon_\gamma (k), \mu(T)] = \left\{1 + \tet{exp}[(\gamma \, \hbar v_F k - \mu)/(k_B T)] \right\}^{-1}$ are Fermi-Dirac distribution functions and the wave function overlaps $\mbb{O}_{\gamma_1,\gamma_2}(\beta_{{\bf k},{\bf k}+{\bf q}})$ are given by \,\cite{huang2019interplay}

\begin{eqnarray}
&& \mbb{O}_{\pm 1,1}(\phi \, \vert \, \beta_{{\bf k},{\bf k}'}) =
\mbb{O}_{1, \pm  1}(\phi \, \vert \, \beta_{{\bf k},{\bf k}'}) =
\frac{1}{4} \, \left(
1  \pm \cos \beta_{{\bf k}, {\bf k}'}
\right)^2  + \frac{1}{4} \cos^2 (2 \phi) \, \sin^2 \beta_{{\bf k}, {\bf k}'} \, , \\
\nonumber 
&&  \mbb{O}_{0, \pm 1}(\phi \, \vert \, \beta_{{\bf k},{\bf k}'}) =
\mbb{O}_{\pm 1,0}(\phi \, \vert \, \beta_{{\bf k},{\bf k}'}) =
 \frac{1}{2} \, 
\sin^2 (2 \phi) \, \sin^2 \beta_{{\bf k}, {\bf k}'} 
\label{OO}
\end{eqnarray}
Interestingly, an intra-band element corresponding to the transitions inside the flat band $0 \leftrightarrow 0$ exists and is equal to 
$\mbb{O}_{\pm 0,0}(\phi \, \vert \, \beta_{{\bf k},{\bf k}'}) = \cos^2 \beta_{{\bf k},{\bf k}'}  + \cos^2 (2 \phi) \sin^2 \beta_{{\bf k},
{\bf k}'} $. However, these two transitions do not contribute to the polarization function because the difference of the energies and the 
corresponding distribution functions on the flat band are always zero disregarding of the temperature     
which corresponds to a very limited effect of the presence of a flat band, close to graphene.

\begin{figure}
\centering
\includegraphics[width=0.5\linewidth]{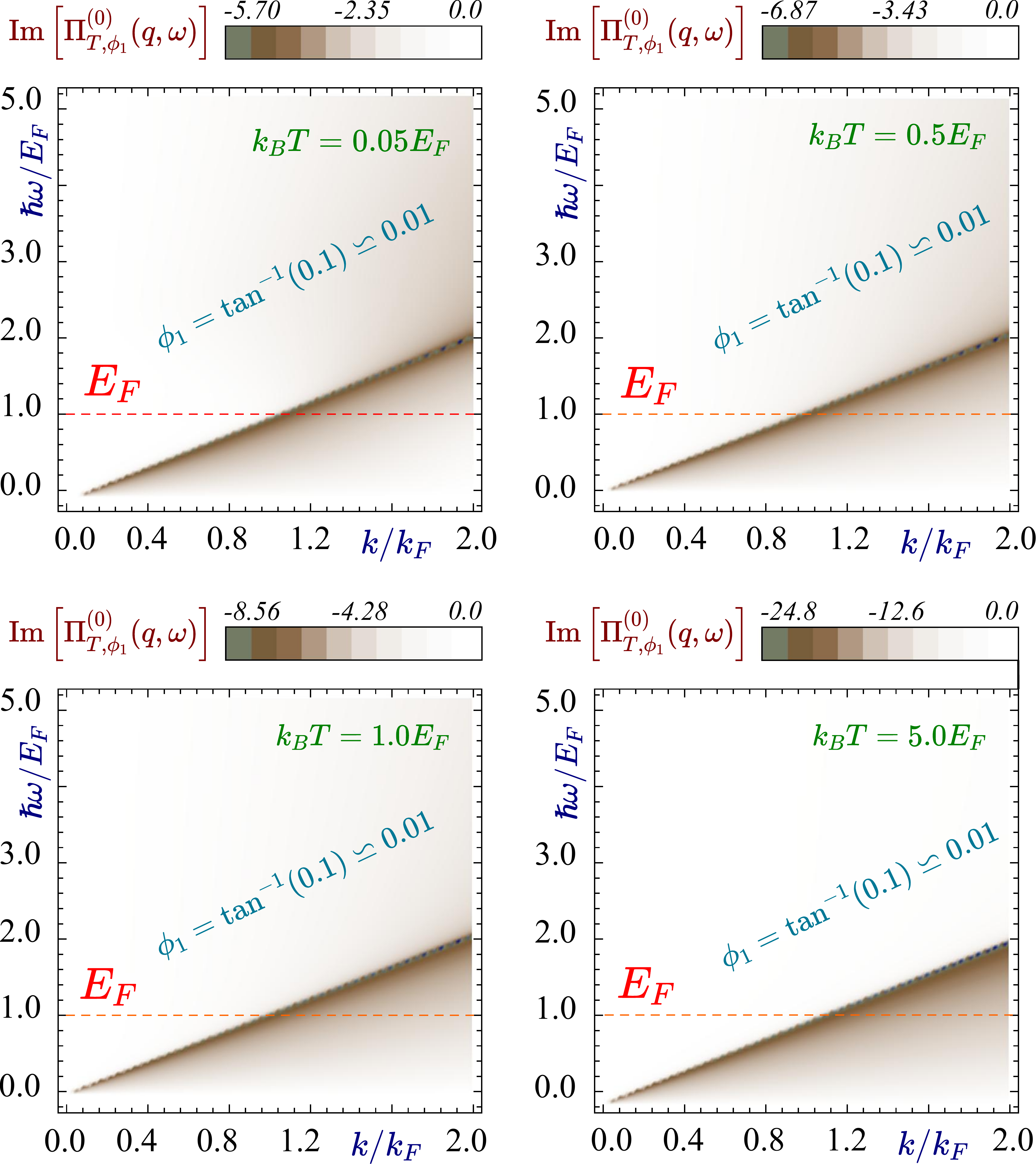}
\caption{(Color online) Particle-hole modes, or single-particle excitation regions
for an $\alpha-\mc{T}_3$ material with a small phase $\phi_1 = \tan^{-1} (0.1) \backsim
0.01$. The particle-hole modes are obtained as the regions of non-zero $\text{Im} \, \Pi^{(0)}_{\phi, T}(q, \omega)$
so that each panel corresponds to a specific temperature ($0.05\,E_F/k_B$, $0.5\,E_F/k_B$ $1.0\,E_F/k_B$ and $5.0\,E_F/k_B$), as labeled. 
Completely white areas with $\text{Im} \, \Pi^{(0)}_{\phi, T}(q, \omega) = 0$ outline the regions with damping-free excitations.}
\label{FIG:1}
\end{figure}

\begin{figure}
\centering
\includegraphics[width=0.5\linewidth]{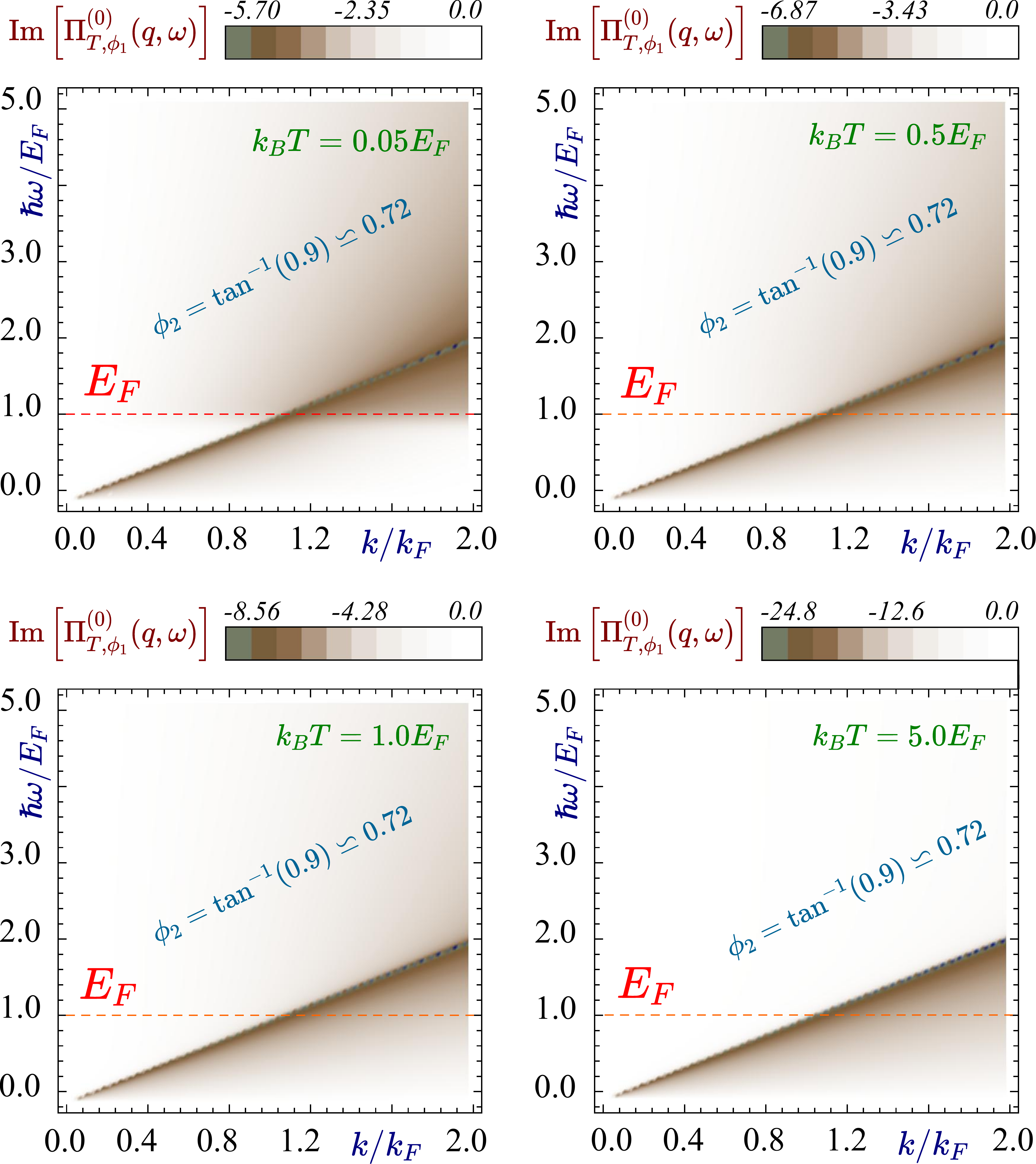}
\caption{(Color online) Particle-hole modes, or single-particle excitation regions
for an $\alpha-\mc{T}_3$ material with a large value of phase $\phi_1 = \tan^{-1} (0.9) \backsim
0.73$. The particle-hole modes are obtained as the regions of non-zero $\text{Im} \, \Pi^{(0)}_{\phi, T}(q, \omega)$
so that each panel corresponds to a specific temperature ($0.05\,E_F/k_B$, $0.5\,E_F/k_B$ $1.0\,E_F/k_B$ and $5.0\,E_F/k_B$), as labeled. 
Completely white areas with $\text{Im} \, \Pi^{(0)}_{\phi, T}(q, \omega) = 0$ outline the regions with damping-free excitations.}
\label{FIG:2}
\end{figure}

\begin{figure}
\centering
\includegraphics[width=0.5\linewidth]{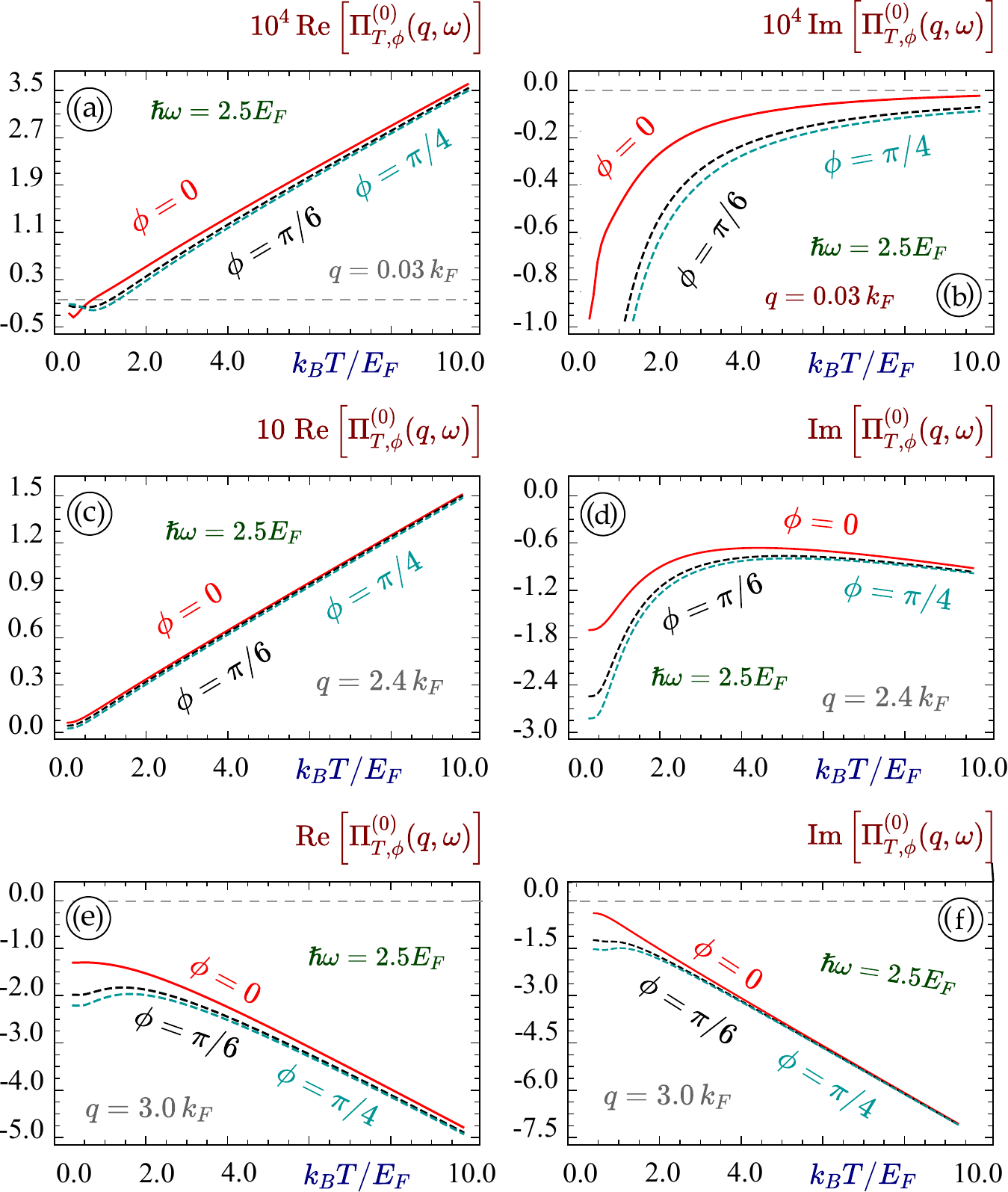}
\caption{(Color online)  Real (left panels) and imaginary (right panels) parts of polarization function as $\Pi^{(0)}_{\phi, T}(q,
\omega)$ a function of temperature for various types of extrinsic (or doped) $\alpha-\mc{T}_3$ materials at specific points with a fixed frequency $\hbar \omega = 2.5\,E_F$ for all panels, and the wave vector $q=0.03\,k_F$ (long-wave limit) for top panels $(a)$ and $(b)$, $q=2.4\,k_F$ for middle plots $(c)$ and $(d)$, and $q=3.0\,k_F$ for the lower panels $(e)$ and $(f)$. In each plot, red, black and blue curves correspond to $\phi= 0$ (graphene), $\phi= \pi/6$ and $\phi= \pi/4$ (a dice lattice), as labeled.}
\label{FIG:3}
\end{figure}

\begin{figure}
\centering
\includegraphics[width=0.5\linewidth]{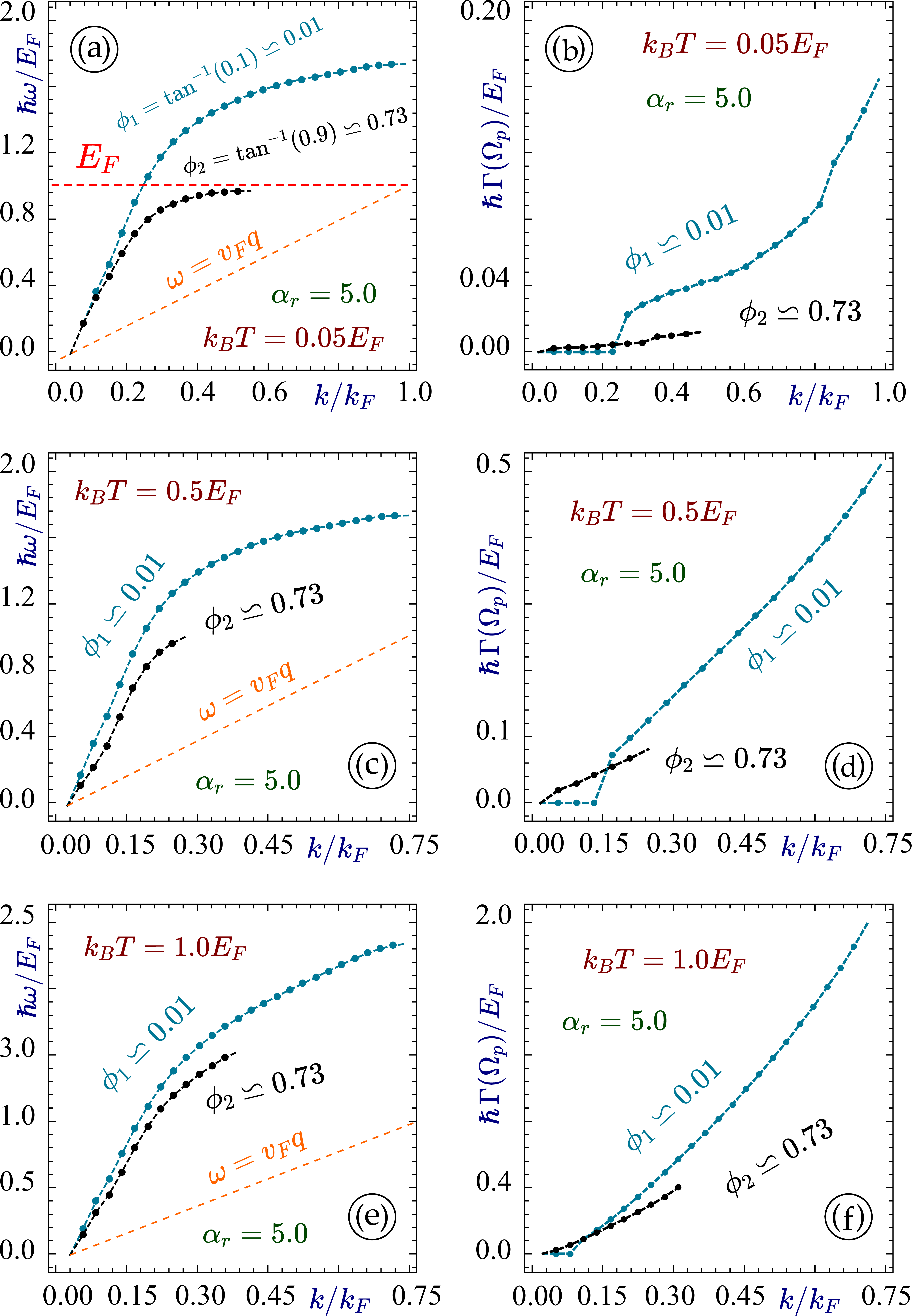}
\caption{(Color online) Plasmon dispersions (left panels) and their damping rates (right panels) for $\alpha-\mc{T}_3$ materials at different temperatures. Top panels $(a)$ and $(b)$ describe the situation for a low temperature $k_B T = 0.05 \, E_F$, middle plots $(c)$ and $(d)$ - for $k_B T = 0.5 \, E_F$, and the lower panels $(e)$ and $(f)$ - for $k_B T = 1.0 \, E_F$. In each panel, a blue curve is related to a material with $\alpha = 0.1$ (nearly graphene), while the black one - to $\alpha = 0.9$ (nearly dice lattice). The relative dielectric constant is taken $\alpha_r = 5.0$ for all cases.}
\label{FIG:4}
\end{figure}

\begin{figure}
\centering
\includegraphics[width=0.5\linewidth]{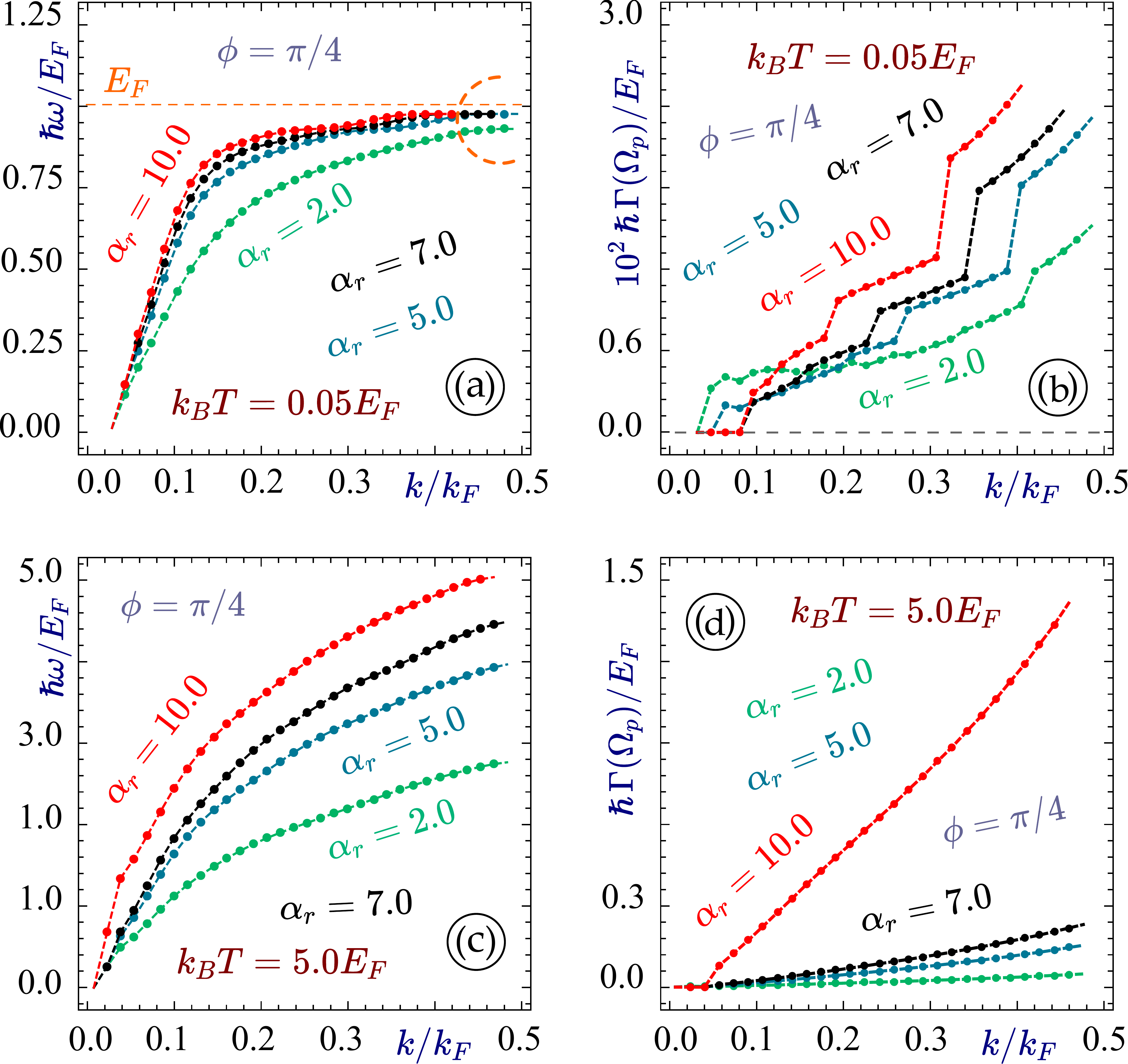}
\caption{(Color online) The suppression of the so-called ``plasmon pinching effect" in a dice lattice at  finite temperature. Panel $(a)$ 
demonstrates the plasmon dispersions for a dice lattice ($\phi = \pi/4$) at or  nearly zero temperature $k_B T = 0.05 \,E_F$ for different 
values of a dielectric constant $\alpha_r = 2.5$, $5.0$, $7.0$ and $10.0$ so that all the branches intersect at $\hbar \omega = \hbar v_F 
q = E_F$. Similar plasmon dispersions at a large temperature $k_B T =5.0 \,E_F$ are shown in plot $(c)$. The right panels 
$(b)$ and $(d)$ represent the corresponding damping rates. 
}
\label{FIG:5}
\end{figure}

\begin{figure}
\centering
\includegraphics[width=0.65\linewidth]{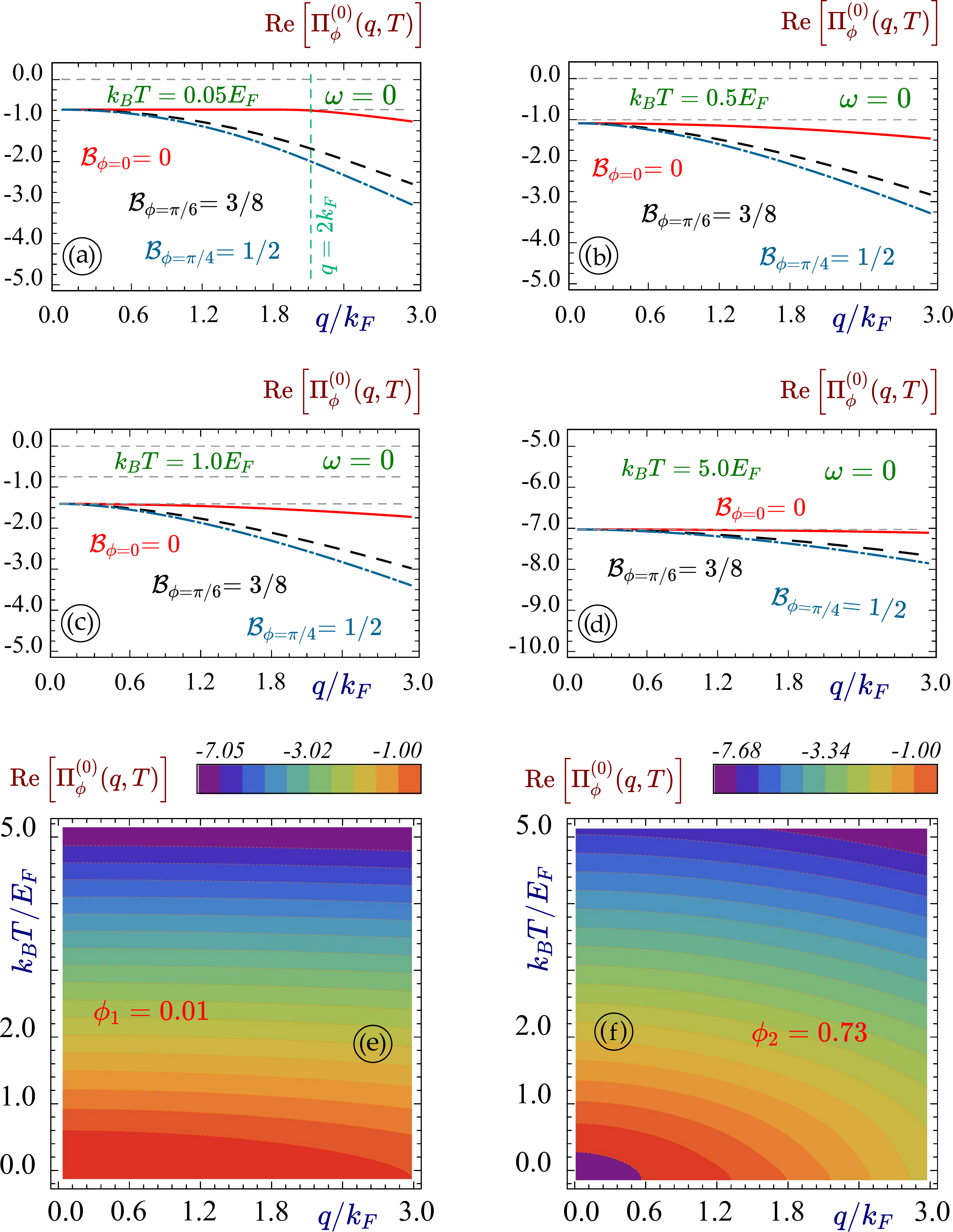}
\caption{(Color online) Static limit of the real part polarization function $\Pi^{(0)}_{\phi, T}(q, \omega = 0 \, \vert \, \mu)$ (its imaginary part is zero) for different $\alpha-\mc{T}_3$ materials at a finite temperatures. Panels $(a)$ - $(d)$ show 
$\Pi^{(0)}_{\phi, T}(q, \omega = 0 \, \vert \, \mu)$ as a function of wavevector $q$ for $T = 0.05\,E_F/k_B$, $0.5\,E_F/k_B$, $1.0\,E_F/k_B$
and $5.0\,E_F/k_B$, correspondingly. In each panel, the red blue and black curves are related to different phases ($\phi = 0$, $\pi/6$ and 
$\pi/4$), as labeled. The two lower panels $(e)$ and $(f)$ demonstrate density plots of the 
$\Pi^{(0)}_{\phi, T}(q, \omega = 0 \, \vert \, \mu)$ as a function of the wave vector $q$ and temperature $T$ simultaneously for phase 
$\phi_2= \tan^{-1}(0.1) = 0.01$ (nearly graphene) and $\phi_2 = \tan^{-1}(0.9) \backsimeq 0.73$ (nearly a dice lattice).}
\label{FIG:6}
\end{figure}
\section{Long wavelength limit of $\Pi_{\phi, T}^{(0)} (q,\omega)$ at zero and finite temperatures}
\label{s03}

In this section, our aim is to find the long wavelength limit ($q \longrightarrow 0$) of the polarization function and the plasmon dispersions which are often used to evaluate a number of crucial properties of the material and the plasmon behavior in various heterostructures\,\cite{gumbs2015nonlocal, iurov2017controlling} Our results for $T = 0$ and $q \ll E_F/(\hbar v_F)$ are based on  analytical expressions for the polarization function demonstrated in  Ref.~[\onlinecite{Malc01}]. Strictly speaking, our results for an arbitrary $\alpha$-T$_3$ case is an interpolation (not a rigorous derivation): we presented the expressions for a dice lattice as the corresponding result for graphene plus additional non-graphene terms and interpolated these extra terms $\backsim \mc{B}_\phi = 1/2 \, \sin^2 (2 \phi)$ for all $\alpha - \mc{T}_3$ materials. However, we strongly believe that this result is precise because exactly the same $\phi-$dependence was rigorously obtained in our thorough static limit derivation in Section \ref{s04}.

\medskip
\par
Following our derivation in  Appendix \ref{AppB}, we write

\begin{equation}
\label{P0}
 16 \pi \hbar^2 g^{-1} \, \Pi_{\phi}^{(0)} (q,\omega) =  2 E_F \left\{
\frac{2}{\omega^2} - \frac{\mc{B}_\phi (\hbar \omega)^2}{\left[ 4 E_F^2 - (\hbar \omega)^2 \right]^2}
\right\} \,q^2 - (1 + 6 \mc{B}_\phi) \, \frac{q^2}{\hbar \omega} \,
\ln \left[
\frac{2 E_F + \hbar \omega}{2 E_F - \hbar \omega}
\right] \, . 
\end{equation}
 Now we are assured that the plasmon branch varies as $\backsim \sqrt{q}$, as it should be for any two-dimensional Dirac cone lattice. This means that for $q \longrightarrow 0$, the plasmon is located at an energy $\Omega_p \backsim \sqrt{q} \ll E_F$ and Eq.~\eqref{AP0} could be further simplified as 

\begin{equation}
\label{P00}
 16 \pi g^{-1} \, \text{Re}\,\Pi_{\phi, T=0} (q,\omega) = \left\{ 
\frac{4 E_F}{(\hbar \omega)^2} - \left(1+ 6 \mc{B}_\phi \right) \frac{2}{E_F} - \mc{B}_\phi \frac{(\hbar \omega)^2}{(2 E_F)^3}  \right\} \, q^2 
\, .
\end{equation}
The analytical expression for the real part of the zero-temperature polarization function could be found only for  finite and sufficiently large doping level $E_F$. The polarization function for finite doping cannot be presented as intrinsic $\text{Re}\,\Pi_{\phi, T=0} (q,\omega)$ plus an additional term proportional to the Fermi energy. Therefore, the integral transformation in Eq.\  \eqref{Pi0T} cannot be applied to the real part of the polarization function \ref{P00} and the analytical expression for the temperature-induced modification of the polarization function needs to  be generalized in the presence of a flat band.  We would, however, estimate the finite-temperature real polarizability using the fact that at $T=0$ K the flat band has only very limited effect on the polarization function (next order of $\hbar \omega$) and the plasmon branch in the long wavelength limit. Apart from that, at high temperature $k_B T \gg E_F$ and the long wavelength limit $q \ll k_F$ the situation is such that the finite electron doping $E_F$ has  a negligible effect on the polarization function and the plasmon branch,\,\cite{sarma2013intrinsic} i.e., at high temperature the extrinsic and intrinsic materials' behavior is almost identical. Therefore, in the lowest-order approximation, we can assume that the polarization function is the same as for undoped graphene.\,\cite{sarma2013intrinsic} 

\begin{equation}
\label{RET}
\text{Re}\,\Pi_{\phi, T \gg E_F} (q,\omega) = \frac{g \ln 2}{2 \pi}\, k_B T \, \frac{q^2}{\omega^2} \, 
\end{equation}

\medskip 

In the lowest order for which a coefficient $\mc{B}_\phi = 1/2 \sin^2 (2 \phi)$ would not vanish, the  zero-temperature plasmon
branch for an $\alpha-\mc{T}_3$ lattice is given as 

\begin{equation}
\Omega_p^2(q, T=0 \, \vert \, E_F, \phi) = \frac{2 g E_F^2 \, q}{4 E_F + g \left(1+ 6 \mc{B}_\phi \right) \, q} 
\backsimeq \frac{g E_F}{2 \hbar^2} \,q - \left(1+ 6 \mc{B}_\phi \right) \frac{g^2}{8 \hbar^2} \, q^2 + ...  \, . 
\end{equation}
The imaginary part of the polarization function for an $\alpha-\mc{T}_3$ model is calculated as (see Appendix \ref{AppB})

\begin{eqnarray}
\label{Immain}
&& \text{Im} \, \Pi^{(0)}_{\phi, T=0}(q, \omega)  \backsimeq - \frac{g \, q^2}{16 \,\hbar \omega} \Big\{
4  \mc{B}_\phi \Theta( \omega - E_F) + (1 - 2 \mc{B}_\phi) \, \Theta \left[ \omega - 2 E_F \right] 
\Big\} \, . 
\end{eqnarray}

\medskip 
\par
The imaginary part of the finite-temperature polarization function in Eq.  \eqref{Immain} could be easily calculated as

\begin{equation}
\text{Im} \, \Pi^{(0)}_{\phi, T \gg T_F}(q, \omega \, \vert \, \mu)  \backsimeq \frac{g}{16}\,\frac{q^2}{\hbar \omega}\,
\left\{
4 \mc{B}_\phi \, \tanh \left( \frac{\omega - \mu}{2 k_B T} \right) + (1 - 2 \mc{B}_\phi) \, 
 \tanh \left[ \frac{\omega - 2 \mu}{4 k_B T} \right] + 
(1 + 2 \mc{B}_\phi) \, 
 \tanh \left[ \frac{\mu}{2 k_B T} \right]
\right\} \, . 
\end{equation}
At  high temperature $k_B T \gg \mu$ or $k_B T \gg \omega$, we can approximate $\tanh x \backsimeq x$ so that 

\begin{equation}
\label{IMT}
\text{Im} \, \Pi^{(0)}_{\phi, T \gg T_F}(q, \omega \, \vert \, \mu)  \backsimeq - \frac{g}{64}\,\frac{q^2}{k_B T}\,
\left(1 + 6 \mc{B}_\phi
\right) \, . 
\end{equation}
The effect of  doping on the imaginary part of the  polarizability is only seen in the next-order corrections:

\begin{equation}
\label{ImPT}
\text{Im} \, \Pi^{(0)}_{\phi, T \gg T_F}(q, \omega \, \vert \, \mu)  \backsimeq - \frac{g}{64}\,\frac{q^2}{k_B T}\,
\left\{1 - \frac{1}{48}\, \left( \frac{\omega}{k_B T} \right)^2 + 6 \mc{B}_\phi \, \left[
1 - \frac{5}{48}\, \left( \frac{\omega}{k_B T} \right)^2 
\right]
\right\} \, . 
\end{equation}
Here, we assumed that at high temperature $k_B T \gg E_F$ we can use $\omega \backsim k_B T \gg \mu(T)$.

\medskip 
The decay rate $\Gamma_{\phi, T}[q,\Omega_p(q) \, \vert \, \mu] $ of the obtained plasmon branches for the case of low damping could be approximated in the first order of $\Gamma(q,\Omega_p)/\Omega_p(q)$ as 

\begin{equation}
\label{Gamma}
\Gamma_{\phi, T}[q,\Omega_p(q) \, \vert \, \mu] = \text{Im} \, \Pi^{(0)}_{\phi, T}(q, \omega \, \vert \, \mu) \, \left\{
\frac{\pr}{\pr \omega} \text{Re} \, \Pi^{(0)}_{\phi, T}(q, \omega \, \vert \, \mu) \Big|_{\omega = \Omega_p(q)}
\right\}^{-1} \, . 
\end{equation}
Such an approximation works particularly well at a large temperature $k_B T \gg E_F$ and a small wave vector $q \ll k_F$ when the  real part grows $\backsim T$ and imaginary part is reduced as $\backsim 1/T$. For the intermediate part, the validity of approximation  \ref{Gamma} is not obvious and must be verified by the effect of the next-order terms $\backsim \Gamma^2$ etc.

\medskip 
\par
Substituting the obtained expressions \eqref{RET} and \eqref{IMT} for the high-temperature limit of the real and imaginary parts of the polarization function into the plasmon equation \eqref{eps01}, we obtained the thermal plasmon dispersions and their damping rates is 

\begin{eqnarray}
\label{TOmG}
&& \Omega_p(q, T) = \sqrt{\alpha_r g \log(2) \, \hbar v_F q \,k_B T} \, , \\
\nonumber  
&& \Gamma_{\phi, T}[q,\Omega_p(q)] = \frac{g \pi}{32} \,\sqrt{\log(2)} \, \frac{(\alpha_r \, \hbar v_F q)^{3/2}}{\sqrt{k_B T}} \, \left(
1+ 6 \mc{B}_\phi
\right) \, .
\end{eqnarray} 
Qualitatively, the behavior of the high-temperature plasmons in $\alpha-\mc{T}_3$ and in a dice lattice is similar to graphene.

\medskip 
\par
The particle-hole mode region, corresponding to finite imaginary part of the polarization function, for different $\alpha-\mc{T}_3$ at finite temperatures, are  presented at Figs.~\ref{FIG:1} and \ref{FIG:2}. We observe that $\text{Im} \, \Pi^{(0)}_{\phi, T}(q, \omega \, \vert \, \mu)$ is increasing with temperature for  sufficiently large values of the wave vector $q$ which is opposite to its behavior for $q \to 0$. In the long wavelength limit, the imaginary part of the polarization function drops exactly as $\backsim 1/T$ which is especially well demonstrated at high temperatures (see Eqs \eqref{TOmG} and \eqref{IMT}). Once the temperature becomes finite, we no longer see a clear division of the areas with zero and non-zero polarization function which define the areas of undamped plasmons and the single-particle excitation spectrum.  The largest values of imaginary polarization is observed along the main diagonal $\omega = v_F q$ for all temperatures.

\medskip
\par
The real part of the polarization function, presented in Fig.~\ref{FIG:3} increases nearly proportional with the temperature. However, it could be either positive or negative depending on its location relative to the main diagonal. Both real and imaginary parts of $\Pi^{(0)}_{\phi, T}(q, \omega \, \vert \, \mu)$ for any $\alpha > 0$ are bigger than for graphene, and their values monotonically increase with increasing phase  $\phi$ due to the fact that if the flat band is present more transitions from and to the flat band have to be taken into account, and the weight of this contributing term keeps monotonically increasing with phase $\phi$.

\medskip
\par
Plasmon damping rates for various, calculations using Eq.~\eqref{Gamma} for $\alpha-\mc{T}_3$ materials and different temperatures are shown in Fig.~\ref{FIG:4}. Generally, the damping rate is affected by phase $\phi$ (the effect of the presence of the flat band) a lot more than the actual plasmon dispersions $\Omega_p(q)$. In the long wavelength limit, we see that the dispersions of the plasmon branches only decrease in the second order of $q$. For larger wave vectors $q$, this decrease becomes more substantial, but the most important difference is between a dice lattice (and any $\alpha-\mc{T}_3$ with $\alpha > 0$) and graphene is the strong damping at low frequencies $\hbar \omega \backsimeq E_F$ which makes the plasmon branch strongly damped at any temperature.

\medskip 
\par
At high temperature, the plasmon branches corresponding to phases $\phi_1$ (almost graphene) and $\phi_2$ (almost a dice lattice) are located closer to each other meaning that the effect of the flat band is largely suppressed, but for a large phase the plasmons become  highly damned at lower frequencies irrespective of the temperature. The decay rate of the plasmons is also strongly affected by the $\omega$- derivative, or how closely the branch is located to the main diagonal $\omega = v_F q$. Thus, we see that for graphene-like materials the damping is stronger for small wave vector values (black curves in Figure \ref{FIG:4}. 
 
\medskip
\par
Additionally, it seems interesting to investigate the so-called plasmon pinching effect meaning that all the plasmon dispersions corresponding to different relative dielectric constants $\alpha_r$ intersect at $\hbar \omega = \hbar v_F q = E_F$. We see that at finite temperature this effect no longer exists, and the plasma branches are well separated, as well as the corresponding damping rates. The plasmon branches for the smaller values of $\alpha_r$ are located at lower frequencies closer to the main diagonal which increases the $\omega$-derivative of the real part of the polarization function.  

\section{Static limit of the polarization function}
\label{s04}

We now turn our attention to the calculation of a static ($\omega = 0$) limit of the polarization function $\Pi_{0,\phi}(q,T)$ for $\alpha-\mc{T}_3$ materials with arbitrary $0 < \alpha < 1$   and finite temperature $T$. The obtained function also referred to as the Lindhard function must be real  for all wave vectors $q$ since $q > \omega  = 0$ and $\omega < \mu(T)$ conditions are always satisfied.  

\medskip  
\par
As we demonstrate in Appendix \ref{AppC}, the zero-frequency electron polarization function  can  be written as

\begin{equation}
\Pi_\phi^{(0)}(q,T) = \frac{g}{2 \pi} \, \frac{1}{(\hbar v_F)^2} \, \left\{ 
\chi^{(+)}_{\phi}(q,T) + \chi^{(0)}_{\phi}(q,T) + \chi^{(-)}_{\phi}(q,T) 
\right\} \, ,
\end{equation}

where

\begin{eqnarray}
&& \, \chi^{(+)}_{\phi}(q,T) = - \mu(T) -  k_B T \, \log\left\{
1 + \tet{exp} \left[
\frac{-\mu}{k_B T}
\right]
\right\} + \hbar v_F 
\int\limits_0^{q/2} \frac{dk \, \left[1 + 2 \mc{B}_\phi -  (2 k / q)^2 \right]}{ \left[
1+ \tet{exp} 
\left\{[\varepsilon(k) - \mu]/(k_B T)
\right\}
\right]
 \, \sqrt{1 - (2 k / q)^2}}    \\
\nonumber
&&  +
2 \hbar v_F \mc{B}_\phi \int\limits_0^{q/2}  \, 
\frac{
dk \, [q/(2 k)]^2
}{
1+ \tet{exp} 
\left\{[\varepsilon(k) - \mu]/(k_B T)
\right\}
}
\, \left\{ 1 - 
\left[1-(2k/q)^2\right]^{-1/2}
\right\} + 2 \hbar v_F \mc{B}_\phi \int\limits_{q/2}^{\infty} dk \, 
\left(\frac{q}{2 k }\right)^2
\,
\left\{
1+\tet{exp}\left[
\frac{\hbar v_F k - \mu}{k_B T}
\right] 
\right\}^{-1} \,  ,
\end{eqnarray}

\begin{eqnarray}
\nonumber
&& \, \chi^{(-)}_{\phi}(q,T) = -(1 + 4 \mc{B}_\phi) \frac{\pi}{8} \, \hbar v_F q - k_B T \, \log\left\{
1 + \tet{exp} \left[
\frac{-\mu}{k_B T}
\right]
\right\} + \hbar v_F 
\int\limits_0^{q/2} \frac{dk \, \left[1 +  2 \mc{B}_\phi -  (2 k / q)^2 \right]}{ \left[
1+ \tet{exp} 
\left\{[\varepsilon(k) + \mu]/(k_B T)
\right\}
\right]
 \, \sqrt{1 - (2 k / q)^2}}   \\
\nonumber
&&  +
2 \hbar v_F \mc{B}_\phi \int\limits_0^{q/2}  \, 
\frac{
dk \, [q/(2 k)]^2
}{
1+ \tet{exp} 
\left\{[\varepsilon(k) + \mu]/(k_B T)
\right\}
}
\, \left\{ 1 - 
\left[1-(2k/q)^2\right]^{-1/2}
\right\} + 2 \hbar v_F \mc{B}_\phi \int\limits_{q/2}^{\infty} dk \, 
\left(\frac{q}{2 k }\right)^2
\,
\left\{
1+\tet{exp}\left[
\frac{\hbar v_F k + \mu}{k_B T}
\right] 
\right\}^{-1} \,  ,
\end{eqnarray}
and 

\begin{equation}
\chi^{(0)}_{\phi}(q,T) = 0 \, 
\end{equation}
represent three contributing terms with the conduction, valence and flat band types of the Fermi-Dirac distribution function 
$f(\gamma,k) = \left\{ 1+ \tet{exp} [(\gamma \hbar v_F k - \mu)/(k_B T)]\right\}^{-1}$, where $\gamma = \pm 1$ and $0$.

\medskip 
At zero temperature, an exact analytical formula could be obtained

\begin{eqnarray}
&& \, \chi^{(+)}_{\phi}(q,T \rightarrow 0 ) = - E_F + (1 + 4 \mc{B}_\phi) \frac{\pi}{8} \, \hbar v_F q - \mc{B}_\phi \frac{(\hbar v_F q)^2}{2 E_F} \,  ,
\end{eqnarray} 
for $q < 2 k_F$ and

\begin{eqnarray}
\chi^{(+)}_{\phi}(q,T \rightarrow 0) = && - E_F + \frac{E_F}{2} \sqrt{1 - \left( \frac{2 k_F}{q}\right)^2} + \hbar v_F q \, \left(\frac{1}{4} + \mc{B}_\phi
\right) \, \arcsin \left( 
\frac{2 k_F}{q} \right)   \\
\nonumber 
&& - \mc{B}_\phi \frac{(\hbar v_F q)^2}{2 E_F} \left[ 1 - \sqrt{1 - \left( \frac{2 k_F}{q}\right)^2} \, \right]
 \,  
\end{eqnarray}  
for $q > 2 k_F$. Thus, $\chi^{(+)}_{\phi}(q,T)$ could be presented by a single unified expression 

\begin{eqnarray}
\chi^{(+)}_{\phi}(q,T \rightarrow 0 ) = &&  - E_F + (1 + 4 \mc{B}_\phi) \frac{\pi}{8} \, \hbar v_F q - \mc{B}_\phi \frac{(\hbar v_F q)^2}{2 E_F}  +
\left\{ 
\frac{1}{2}\,\left(E_F + \mc{B}_\phi \, \frac{q^2}{E_F}  \right) \sqrt{1 - \left( \frac{2 k_F}{q}\right)^2}  \right. \\
\nonumber
&& - \left. \hbar v_F q \, \left[\frac{1}{4} + \mc{B}_\phi
\right] \, \arccos \left( 
\frac{2 k_F}{q} \right) \, 
\right\} \, \Theta(q - 2 k_F)
 \,  .
\end{eqnarray} 

The hole contribution term $\chi^{(-)}_{\phi}(q,T)$ at $T \rightarrow 0$ becomes

\begin{eqnarray}
&& \, \chi^{(-)}_{\phi}(q,T \rightarrow 0) = -(1 + 4 \mc{B}_\phi) \frac{\pi}{8} \, \hbar v_F q \,  ,
\end{eqnarray}
which for a dice lattice with $\mc{B}_\phi = 1/2$ exactly matches the corresponding results of Ref.~[\onlinecite{Malc01}].

\medskip 
\par
We plot the results for the finite-temperature static polarizability in Fig.~\ref{FIG:6}. The main difference between graphene and $\alpha-\mc{T}_3$ is a smooth variation of the Lindhard function around $q = 2 k_F$ for all $\alpha \neq 0$ which means a lower order for the Thomas-Fermi decay not only for a dice lattice but for all other $\alpha-\mc{T}_3$ lattices.  Apart from that, the $q$-dependence of the static polarization function is smooth for any non-zero temperature regardless of of the phase $\phi$. 

\medskip 
\par
We have observed that the Lindhard function for any $\alpha > 0$ exceeds that for graphene and it always increases with the phase $\phi$ for both zero and finite temperatures since for a finite $\phi$ more electron transitions with a larger weight contribute to the sum in Eq.\ \eqref{Pi0T}. There is also a substantial increase of $\Pi_\phi^{(0)}(q,T)$ with rising temperature which is a lot more noticeable than its $q-$ dependence. The finite-temperature static dielectric function $\epsilon(q,T)$ is the central quantity for calculating the Boltzmann conductivity in a wide class of pseudospin-1 Dirac materials.

\section{Concluding  remarks and Summary}
\label{s05}

In this work, we have investigated the polarization function, plasmons and their damping rates in $\alpha-\mc{T}_3$ materials. The calculations were carried out at finite temperature with the main focus on the temperature dependence of the effect of the presence of a flat band in the electron bandstructure. 

\medskip 
\par
The particle-hole modes, determined by the finite imaginary part of the polarization function, demonstrate completely different behaviors dependent on their location in the $(q,\omega)$-plane. In the long wavelength limit, the high-temperature behavior of the plasmon and its damping is qualitatively similar to graphene, i.e.,  $\text{Im} \, \Pi^{(0)}_{\phi, T \gg T_F}(q, \omega \, \vert \, \mu)$ falls off on $1/T$.  On the other hand, the real part of the polarization function increases as $\backsim T$ so that the plasmon is well defined as long as $k_B T /\gg E_F$.  However, the actual values of $\text{Im} \, \Pi^{(0)}_{\phi, T \gg T_F}(q, \omega \, \vert \, \mu)$ and $\Gamma_{\phi, T}[q,\Omega_p(q)]$ differing substantially from that in graphene being significantly larger at any finite temperature. 
 
\medskip 
\par
We have obtained a closed-form analytical expression for the long wavelength limit of the polarization function and plasmon dispersions for all $\alpha-\mc{T}_3$ materials.  Additionally, we have demonstrated that this long wavelength behavior is nearly similar to the case of graphene $(\phi = 0)$ and the effect of the flat band is only revealed in the next order in $q$ as $\backsim q^2$. The plasmon pinching effect is suppressed at any finite temperature so that the plasmon branches corresponding to different dielectric constants $\alpha_r$ would no longer intersect at one point.  We have also obtained semi-analytical expressions for the Lindhard function, or the static limit of the polarization function. These results were obtained  for various temperatures with completely analytical approximations for the high, low and zero temperatures for an arbitrary $\alpha-\mc{T}_3$ material with $0 < \alpha < 1$.  

 \medskip
\par 
The static polarization function is of special importance and is is often highlighted for the role it plays in collective phenomena. This is so because the static dielectric function is the key component for the temperature-dependent conductivity in the presence of charged  impurities, finite-temperature screening and, more generally, Boltzmann transport. It is also employed in evaluating the screened potential of a point charge, Friedel oscillations and  the effect of an electric or magnetic impurity on the electron response, as well as the RKKY interaction. Our convenient, handy and ready-to-use semi-analytical expressions are expected to motivate researchers leading to a new wave of  research of the many-body properties of $\alpha-\mc{T}_3$ materials. 

\medskip
\par
In brief, we have discovered some important information on the plasmons and collective behavior of pseudospin-1 Dirac materials at finite temperatures. We are confident that these results will find numerous applications in both fundamental research of these innovative systems and their applications. Specifically, we believe that our crucial analytical expressions will be useful to experimentalists for predicting the behavior of flat-band materials at experimentally accessible high and room temperatures.

\section{Acknowledgments}
AI would like to acknowledge the funding provided by TRADA-52-113 PSC-CUNY Award No. 64076-00 52. DH
would like to acknowledge the financial supports from Air Force Office of Scientific Research (AFOSR). 
GG would like to acknowledge the support from Air Force Research Laboratory (AFRL) through Contract 
$\#$ FA9453-18-1-0100.

\appendix

\section{Finite-temperature polarization function - miscellaneous derivations}
\label{AppA}

\subsection{Finite-temperature polarization function $\Pi^{(0)}_{\phi,T}(q, \omega \, \vert \,  \mu)$}

We first demonstrate that the finite-temperature polarization function $\Pi^{(0)}_{\phi,T}(q, \omega \, \vert \,  \mu)$  could be obtained as a straightforward integral transformation of its zero-temperature counterpart. The validity of Eq.~\eqref{Pi0T} is based on the assumption that all the quantities in Eq.~\eqref{Pi00} except the Fermi-Dirac distribution  functions $n_F[\gamma \epsilon(k),T]$ are temperature independent and, therefore, the main idea of our derivation is to prove that  $n_F[\gamma \epsilon(k),T]$ is obtained from its zero-temperature counterpart $n_F[\gamma \epsilon(k),T= 0] = \Theta [E_F - \gamma  \epsilon(k)]$ by applying the exactly identical convolution in Eq.\ \eqref{Pi0T} \,\cite{maldague1978many,iurov2016plasmon,iurov2020thermal}

\begin{equation}
\label{i1}
n_F[\epsilon(k),T] = \int\limits_{-\infty}^{\infty} d \eta \, \frac{\Theta [\nu - \epsilon(k)]}{4 k_B T}  \cosh^{-2} \left\{
\frac{\eta - \mu(T)}{2 k_B T}
\right\} = 
\int\limits_{\gamma \epsilon(k)}^{\infty} \frac{d \eta}{4 k_B T}  \cosh^{-2} \left\{
\frac{\nu - \mu(T)}{2 k_B T}
\right\}
\, .
\end{equation}
Integral \eqref{i1} could be easily carried out yielding 

\begin{equation}
\label{i2}
\int\limits_{\gamma \epsilon(k)}^{\infty} \frac{d \eta}{4 k_B T}  \cosh^{-2} \left[
\frac{\eta - \mu(T)}{2 k_B T}
\right] = \frac{1}{2} \, \tanh \left[ \frac{\eta - \mu(T)}{2 k_B T} \right] \Bigg|_{\gamma \epsilon(k)}^{\infty} = 
\frac{1}{2} \left\{
1 - \tanh \left[
\frac{\epsilon(k) - \mu(T)}{2 k_B T}
\right]
\right\}
\, .
\end{equation}
On the other hand, the  Fermi-Dirac function could be presented as 

\begin{equation}
n_F(\xi) = - \tanh \left( \frac{\xi}{2} \right) + n_F(-\xi) 
\end{equation}
or 
\begin{equation}
\label{AnF}
2 n_F(\xi) = - \tanh \left( \frac{\xi}{2} \right) + 1 \, ,
\end{equation}
where $\xi = [\epsilon(k) - \mu(T)]/(2 k_B T)$. Substituting expression \eqref{AnF} into Eq.~\eqref{i2} immediately verifies our statement. 

\begin{figure}
\centering
\includegraphics[width=0.3\linewidth]{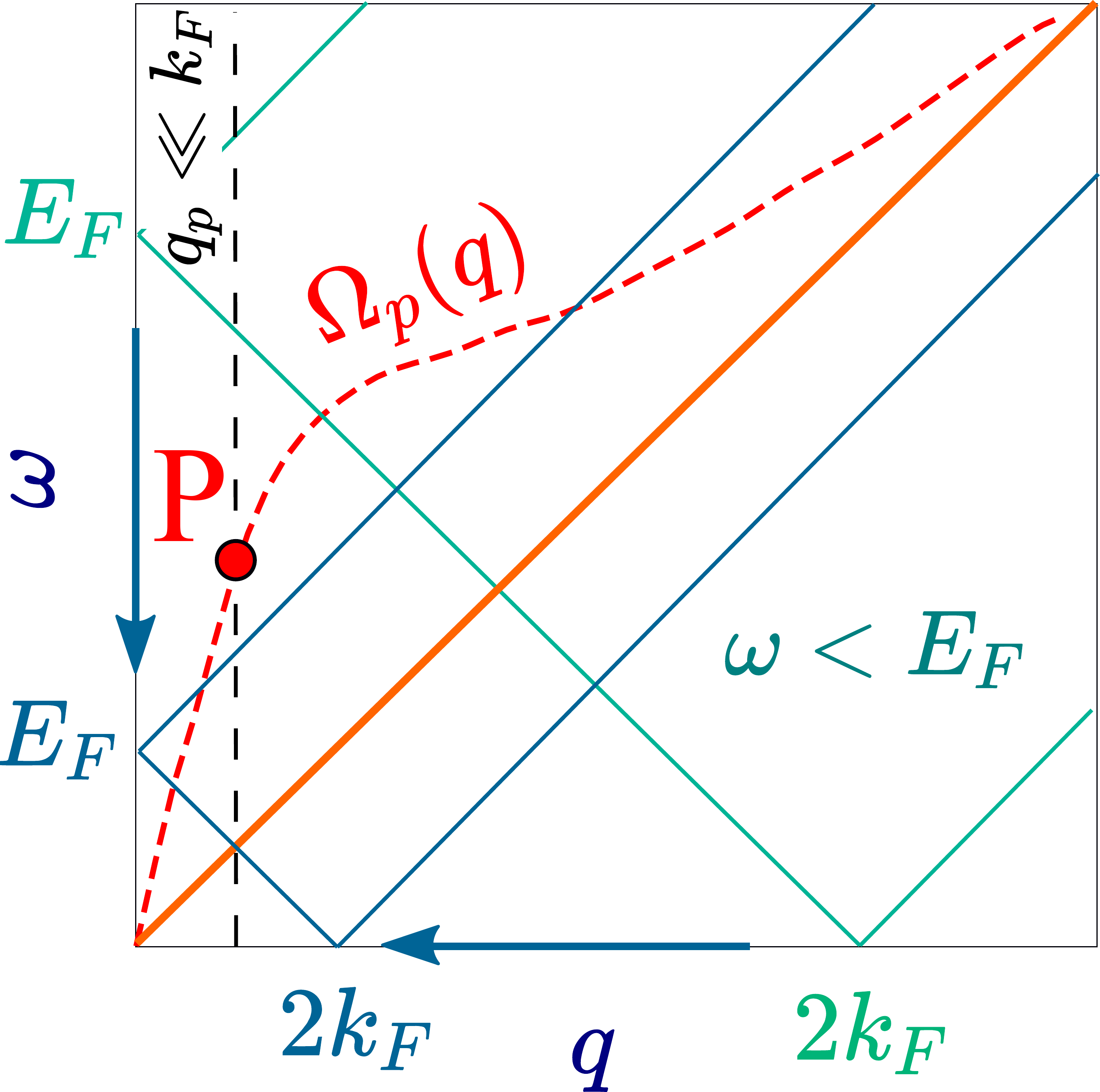}
\caption{(Color online) Illustration for calculating the finite-temperature polarization function $\Pi^{(0)}_{\phi, T}[q, \omega \, \vert \,  \mu(T)]$ by Eq.~\eqref{Pi0T} as an integral over a variable chemical potential $\eta$.}
\label{FIG:1A}
\end{figure}

\par 
For the hole states with $\gamma = -1$ and $\gamma \epsilon(k) < 0$, the Fermi-Dirac distribution functions are presented as  
$n_F[- \epsilon(k) - \mu(T)] = 1 - n_F[\epsilon(k) + \mu(T)]$ and a similar type of transformations could be performed. The undoped 
part of the polarization function does not depend on temperature: 

\begin{equation}
\Pi^{(0)}_{\phi,T}(q, \omega \, \vert \,  \mu = 0) = \frac{1}{4 k_B T} \int\limits_{-\infty}^{\infty}
\frac{d \eta}{\pm [\epsilon(k) + \epsilon(k')] + \hbar \left( \omega + i 0^+ \right) } \, \cosh^{-2} \left\{
\frac{\eta - 0}{2 k_B T}
\right\}  = \Pi^{(0)}_{\phi,T=0}(q, \omega \, \vert \,  E_F = 0) \, ,
\end{equation}
i.e., the finite-temperature undoped polarizability is obtained by the same integral transformation Eq.~\eqref{Pi0T}, and exactly the 
same reasoning could be applied to the transitions from and to the flat band with $\gamma = 0$ and $\epsilon(k) \equiv 0$.

\section{Long wavelength limit $q \longrightarrow 0$ for the polarization function}
\label{AppB}
\subsection{Real part of the polarization function $\text{Im} \, \Pi^{(0)}_{\phi, T=0}(q, \omega)$}

In the region of interest ($R_2$ and $R_4$ defined in Fig.~\ref{FIG:B1}), the polarization function for a dice lattice is obtained as 
\,\cite{Malc01}

\begin{equation}
- \Pi_{\phi}^{(0)} (q,\omega) = \frac{g}{2 \pi} \, E_F + \frac{g}{16 \pi} \, \left[\omega + \frac{q^2}{\omega} \right] \ln X \big|_{X=r_1}^{X=s_1}  + \frac{g}{16 \pi} \, \big[\mbb{G}^{(0)}_q(X)+\mbb{D}^{(\phi)}_q(X)\big] \big|_{X=r}^{X=s} \, ,
\end{equation}
where $\{s,r\} = 2 E_F \pm \omega$ and $\{s_1,r_1\} = E_F \pm \omega$. The two functions

\begin{eqnarray}
&& \mbb{G}^{(0)}_q (X) = - X \sqrt{\frac{X^2 - q^2}{\omega^2 - q^2}} + \frac{q^2}{\sqrt{\omega^2 - q^2}} \, \ln \left[
(X + \sqrt{X^2 - q^2}\,)/q
\right] \, , \\
\nonumber
&& \mbb{D}^{(\phi)}_q (X) = - 2 \sqrt{\omega^2 - q^2} \,\ln \left[
(X + \sqrt{X^2 - q^2}\,)/q
\right] + \left[\omega + \frac{q^2}{\omega} \right] \, \ln \left\{ 
\frac{\omega \sqrt{X^2 - q^2} + (\omega \leftrightarrow X)}{\omega \sqrt{X^2 - q^2} - (\omega \leftrightarrow X)}
\right\}
\end{eqnarray}
correspond to graphene and additional terms obtained for a dice lattice. Here, $(\omega \leftrightarrow X)$ means double variable replacement 
which results in $X \sqrt{\omega^2 - q^2}$ term. 

For a general $\alpha - \mc{T}_3$ model, the equation could be  extrapolated as 

\begin{equation}
- \Pi_{\phi}^{(0)} (q,\omega) = \frac{g}{2 \pi} \, E_F + 2 \mc{B}_\phi \, \frac{g}{16 \pi} \, \left[\omega + \frac{q^2}{\omega} \right] \ln X \big|_{X=r_1}^{X=s_1} + \frac{g}{16 \pi} \, \big[\mbb{G}_q(X)+ 2 \mc{B}_\phi \, \mbb{F}_q(X)\big] \big|_{X=r}^{X=s} \, ,
\end{equation}

\medskip 
\par
At this point, we are in a position to obtain the long wavelength approximation of the polarization function. Calculation shows that 

\begin{eqnarray}
 \mbb{G}^{(0)}_{q \longrightarrow 0} (s) - \mbb{G}^{(0)}_{q \longrightarrow 0} (r)  \backsimeq &&
- 8 E_F  - \frac{4 E_F}{\omega^2} \,q^2 + \frac{q^2}{\omega} \left\{
\ln \left[
\frac{2 E_F + \omega}{2 E_F - \omega}
\right] + \frac{q^2}{2}\,\frac{\omega E_F}{\left( 4 E_F^2 - \omega^2 \right)^2}
\right\}  \, \\
\nonumber 
\backsimeq && 
- 8 E_F - \frac{q^2}{\omega} \left\{ 
\frac{4 E_F}{\omega} - \ln \left[
\frac{2 E_F + \omega}{2 E_F - \omega}
\right]
\right\} + ... \, . 
\end{eqnarray}

\begin{eqnarray}
 \mbb{D}^{(\phi)}_{q \longrightarrow 0} (s) - \mbb{D}^{(\phi)}_{q \longrightarrow 0} (r)  \backsimeq &&
- \left( 2 \omega - \frac{q^2}{\omega} \right) \left\{ 
\ln \left[
\frac{2 E_F + \omega}{2 E_F - \omega}
\right] + \frac{q^2}{2}\,\frac{\omega E_F}{\left( 4 E_F^2 - \omega^2 \right)^2}
\right\}   \\
\nonumber
&& + \left( \omega + \frac{q^2}{\omega} \right) \left\{
\ln \left[
\frac{(2 E_F + \omega)^2(E_F - \omega)}{(2 E_F - \omega)^2(E_F + \omega)}
\right] + \frac{2 E_F \omega\, q^2}{(4 E_F^2-\omega^2)^2} 
\right\} + ... \backsim
\\
\nonumber
&& \backsimeq \omega \ln \left[
\frac{E_F - \omega}{E_F + \omega}
\right] + \frac{\omega^2 E_F \,q^2}{\left( 4 E_F^2 - \omega^2 \right)^2} + \frac{q^2}{\omega} \, 
\ln \left[
\frac{(2 E_F + \omega)^3(E_F - \omega)}{(2 E_F - \omega)^3(E_F + \omega)}
\right] + ... \, . 
\end{eqnarray}

Consequently, the polarization function for the long-wave limit $q \rightarrow 0$ is given by 

\begin{equation}
\label{AP0}
 16 \pi \hbar^2 g^{-1} \, \Pi_{\phi}^{(0)} (q,\omega) =  2 E_F \left\{
\frac{2}{\omega^2} - \frac{\mc{B}_\phi \omega^2}{\left( 4 E_F^2 - \omega^2 \right)^2}
\right\} \,q^2 + (1 + 6 \mc{B}_\phi) \, \frac{q^2}{\omega} \,
\ln \left[
\frac{2 E_F - \omega}{2 E_F + \omega}
\right] \, . 
\end{equation}

Including all the relevant constants, we rewrite the final expression as 

\begin{equation}
\label{AP02}
 16 \pi \hbar^2 g^{-1} \, \Pi_{\phi}^{(0)} (q,\omega) =  2 E_F \left\{
\frac{2}{\omega^2} - \frac{\mc{B}_\phi (\hbar \omega)^2}{\left[ 4 E_F^2 - (\hbar \omega)^2 \right]^2}
\right\} \,q^2 - (1 + 6 \mc{B}_\phi) \, \frac{q^2}{\hbar \omega} \,
\ln \left[
\frac{2 E_F + \hbar \omega}{2 E_F - \hbar \omega}
\right] \, . 
\end{equation}

\par 
We are now assured that the plasmon dispersion varies as $\backsim \sqrt{q}$, as it should be for any two-dimensional Dirac cone lattice. This means that for $q \longrightarrow 0$, the plasmon is located at an energy $\Omega_p \backsim \sqrt{q} \ll E_F$ and Eq.~\eqref{AP0} could be further simplified as 

\begin{equation}
\label{AP00}
 16 \pi g^{-1} \, \Pi_{\phi}^{(0)} (q,\omega) = \left\{ 
\frac{4 E_F}{(\hbar \omega)^2} - \left(1+ 6 \mc{B}_\phi \right) \frac{2}{E_F} - \mc{B}_\phi \frac{(\hbar \omega)^2}{(2 E_F)^3}  \right\} \, q^2 
\, .
\end{equation}
Obviously, this results is valid only for a finite and sufficiently large doping level $E_F$.

\medskip
\subsection{Imaginary part of the polarization function $\text{Im} \, \Pi^{(0)}_{\phi, T=0}(q, \omega)$}

\begin{figure}
\centering
\includegraphics[width=0.3\linewidth]{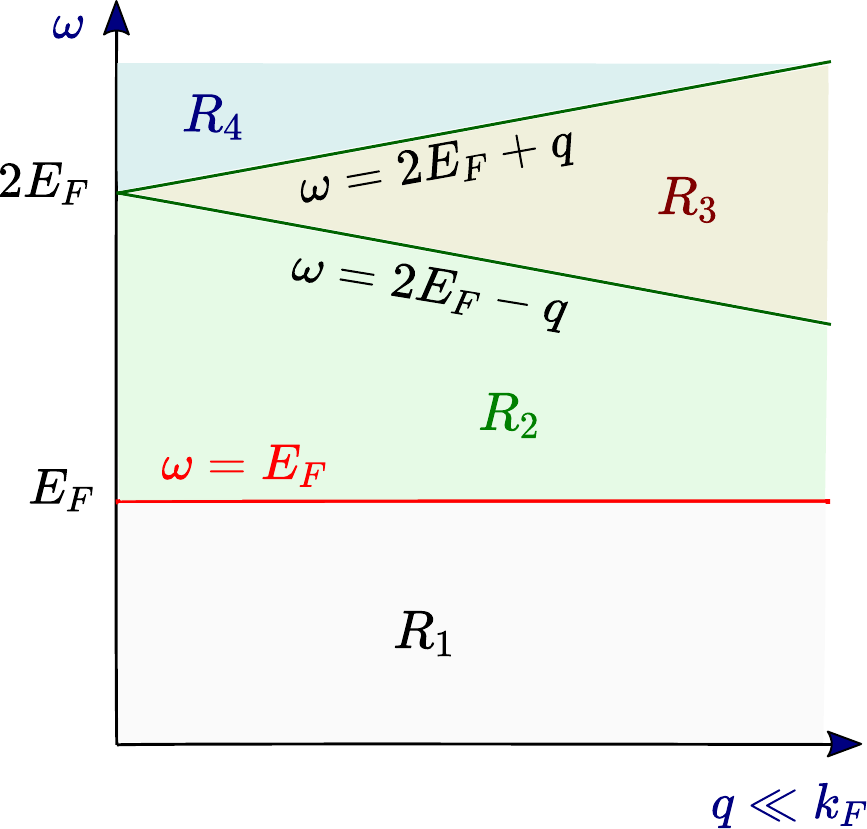}
\caption{(Color online) Schematic graph for the regions with different expressions for the polarization function for $q \longrightarrow 0$.}
\label{FIG:B1}
\end{figure}

We begin our consideration with the imaginary part of the polarization function for a dice lattice for $q \longrightarrow 0$, i,e, 
in regions $R_1$, $R_2$ and $R_4$ assigned in  Fig.~\ref{FIG:B1}:    

\begin{eqnarray}
&& \text{R}_1 = \Theta(E_F - \omega) \, , \\
\nonumber
&& \text{R}_2 = \Theta( \omega - E_F) \, \Theta \left( E_F  - \frac{\omega + q}{2} \right) \, , \\
\nonumber
&& \text{R}_3 =  \Theta \left(  q - \vert \omega - 2 E_F  \vert \right)
\,\, \text{or} \,\, 2 E_F - q < \omega > 2 E_F + q
 , \\
\nonumber
&& \text{R}_4 = \Theta \left(  \frac{\omega - q}{2} - E_F \right) \,\, \text{or} \,\, \omega > 2 E_F + q \, .
\end{eqnarray}

We exclude a small region $R_3$  According to Ref.~[\onlinecite{Malc01}], the polarization function in the regions of interest is given as 

\begin{equation}
\label{Im01}
- \text{Im} \, \Pi^{(0)}_{\phi, T=0}(q, \omega)   = \frac{q^2}{2 \omega} \, \Theta( \omega - E_F) + \Theta \left(  \frac{\omega - q}{2} - E_F \right) 
\left[
\frac{q^2}{4 \sqrt{\omega^2 - q^2}} - \frac{1}{2} \sqrt{\omega^2 - q^2} + \frac{\omega^2 - q^2}{2 \omega}
\right]
\end{equation}
In the long wavelength  limit $q \longrightarrow 0$, Eq.~\eqref{Im01} could be approximated as 

\begin{equation}
- \text{Im} \, \Pi^{(0)}_{\phi, T=0}(q, \omega)   = \frac{q^2}{2 \omega} \, \Theta( \omega - E_F) + \Theta \left(  \frac{\omega - q}{2} - E_F \right) 
\frac{3 q^4}{16 \omega^3}
\end{equation}

For a general $\alpha-\mc{T}_3$ lattice, we can only interpolate the known results for graphene and a dice lattice using a coefficient
$\mc{B}_\phi = 1/2 \sin^2 (2 \phi)$

\begin{equation}
- \text{Im} \, \Pi^{(0)}_{\phi, T=0}(q, \omega)  = \mc{B}_\phi \,\frac{q^2}{\omega} \, \Theta( \omega - E_F) + \Theta \left(  \frac{\omega - q}{2} - E_F \right) 
\left[
\frac{q^2}{4 \sqrt{\omega^2 - q^2}} - \frac{\mc{B}_\phi}{2} \sqrt{\omega^2 -  q^2} + 2 \mc{B}_\phi \frac{\omega^2 - q^2}{2 \omega}
\right]
\end{equation}
Finally, in the long wavelength limit we obtain  

\begin{eqnarray}
\nonumber 
&& \text{Im} \, \Pi^{(0)}_{\phi, T=0}(q, \omega) = - \mc{B}_\phi \, \frac{(\hbar v_F q)^2}{\omega} \, \Theta( \omega - E_F) + \Theta \left[  \frac{\omega - q}{2} - E_F \right] 
\left\{
\frac{1 - 2 \mc{B}_\phi}{4 \omega} (\hbar v_F q)^2 +
\frac{1 + \mc{B}_\phi}{16 \omega^3} (\hbar v_F q)^4 + ...
\right\} \backsimeq \\
\label{AImmain}
&& \backsimeq - \frac{g \, q^2}{16 \,\hbar \omega} \left\{
4  \mc{B}_\phi \Theta( \omega - E_F) + (1 - 2 \mc{B}_\phi) \, \Theta \left[  \frac{\omega - q}{2} - E_F \right] 
\right\} \, . 
\end{eqnarray}
We did not include a narrow triangular region $R_3$ where $ 2 E_F - q < \omega > 2 E_F + q$ into our consideration since for the finite temperature integral $\int\limits_{(\omega-q)/2}^{(\omega+q)/2} \,d \eta$ cannot affect the leading $\backsim q^2$ order of the finite-temperature polarization function. Similarly, we can make a modification $\Theta \left[\omega - q - 2 E_F \right] \rightarrow 
\Theta \left[\omega - 2 E_F \right]$ in Eq.~\eqref{AImmain}.

\section{Lindhard function, static limit of the polarization function}
\label{AppC}

The purpose of the present Appendix is to present a detailed derivation of the static limit of the polarization function  $\Pi_0(q, \omega = 0 \, \vert \, \phi, \mu(T))$ for an $\alpha-\mc{T}_3$ material with any $0 < \phi < \pi/4$ at an arbitrary finite  temperature.   For the static limit $\omega = 0$, Eq.~\eqref{Pi00} is reduced to 

\begin{equation}
\Pi_0(q, T) = \frac{g}{4 \pi^2} \, \int d^2{\bf k} \sum\limits_{\gamma,\gamma' = 0 \, \pm 1} \, 
\mbb{O}_{\gamma_1,\gamma_2}(\beta_{{\bf k},{\bf k}'}) \, \frac{n_F[\gamma, \epsilon(k)] - n_F[\gamma', \epsilon (\vert {\bf k}' \vert )]
}{\epsilon_\gamma (k) - \epsilon_{\gamma'} (\vert {\bf k} + \bf{q} \vert)} \, ,
\label{Pi00T}
\end{equation}
where ${\bf k}' = {\bf k}+{\bf q}$, $\epsilon_\gamma (k) = \gamma \epsilon(k) = \gamma \, \hbar v_F k$, 
$n_F[\gamma, \epsilon(k)] = n_F[\epsilon_\gamma (k), \mu] = \left\{1 + \tet{exp}[(\gamma \, \hbar v_F k - \mu)/(k_B T)] \right\}^{-1}$
is a short-hand notation for the distribution functions and $\mbb{O}_{\gamma_1,\gamma_2}(\beta_{{\bf k},{\bf k}'})$ are the wave function overlaps given by Eq.~\eqref{OO}.

\medskip 

Summation over $\gamma$ and $\gamma' = -1,0,1$ results in 

\begin{equation}
\Pi_{0}(q,T) = \frac{g}{4 \pi^2} \, \frac{1}{\hbar v_F} \int d^2 {\bf k} \left[ 
\chi^{(+)}_{\phi}(q,T) + \chi^{(0)}_{\phi}(q,T) + \chi^{(-)}_{\phi}(q,T) 
\right] \, ,
\end{equation}
where  

\begin{eqnarray}
 \chi^{(-)}_{\phi}(q,T) = && - \frac{n_F(-1,k)-n_F(-1,k')}{k - k'} \,\mbb{O}_{-1,-1}(\beta_{{\bf k}, {\bf k}'}) - 
\frac{n_F(-1,k)+n_F(-1,k')}{k + k'} \,\mbb{O}_{-1,1}(\beta_{{\bf k}, {\bf k}'})  \\
\nonumber 
&& - \left[\frac{n_F(-1,k)}{k} + \frac{n_F(-1,k')}{k'} \right] \,\mbb{O}_{0,-1}(\beta_{{\bf k}, {\bf k}'}) \, ,
\end{eqnarray}

\begin{equation}
\xi^{(0)}_{\phi}(q,T) =  \frac{f(0,k)}{k'} \,\left[\mbb{O}_{0,-1}(\beta_{{\bf k}, {\bf k}'}) - \mbb{O}_{0,1}(\beta_{{\bf k},
{\bf k}'}) \, \right] + 
\frac{f(0,k')}{k} \,\left[\mbb{O}_{-1,0}(\beta_{{\bf k}, {\bf k}'}) - \mbb{O}_{1,0}(\beta_{{\bf k}, {\bf k}'}) \right] = 0
\end{equation}
and 

\begin{eqnarray}
\chi^{(+)}_{\phi}(q,T) = && \frac{f(1,k)+f(+1,k')}{k + k'} \,\mbb{O}_{-1,1}(\beta_{{\bf k}, {\bf k}'}) +
\frac{f(1,k)-f(1,k')}{k - k'} \,\mbb{O}_{1,1}(\beta_{{\bf k}, {\bf k}'})   \\
\nonumber 
&& + \left[\frac{f(1,k)}{k} + \frac{f(1,k')}{k'} \right] \,\mbb{O}_{0,-1}(\beta_{{\bf k}, {\bf k}'}) \, . 
\end{eqnarray}
Here, we have separated the contributions due to the electron $(\gamma = 1)$, hole $(\gamma = -1)$ and flat band $(\gamma = 0)$ states.  We note that for $T=0$ only the terms corresponding to the transition to and from the conduction band   contribute.   The contribution coming from the flat band states $\xi^{(0)}_{\phi}(q,T)$ is equal to zero since $n_F(0,k) = n_F(0,k') = \left[ 1 + \tet{exp}[-   \mu/(k_B T)\right]^{-1}$ is independent of wave vector $k$ or $k^\prime$, and $\mbb{O}_{\pm1,0}(\beta_{{\bf k}, {\bf k}'}) = \mbb{O}_{0, \pm1}(\beta_{{\bf k}, {\bf k}^\prime }) = 1/2 \,\sin^2 (2 \phi) \, \sin^2 (\beta_{\,{\bf k},{\bf k}^\prime})$. 

\medskip
\par 
Also, we will often use the fact that a simultaneous change of variables ${\bf q} \longrightarrow  - {\bf q}$ and ${\bf k} \longrightarrow  -{\bf k}$ does not affect any of the overlaps $\mbb{O}_{\gamma_1,\gamma_2}(\beta_{{\bf k},{\bf k}'})$ since $\beta_{{\bf k},{\bf k}'} = \measuredangle  ({\bf k}, {\bf q}) = \pi - \measuredangle ({\bf k}, -{\bf q})  = \pi - \measuredangle (-{\bf k}, {\bf q})  = \measuredangle ({\bf -k}, {\bf -q})$ so that 

\begin{eqnarray}
&& \frac{k + ( - {\bf q}) \cdot {\bf k}}{k \vert - {\bf q} + {\bf k} \vert} \longrightarrow 
\frac{(-k)^2 + {\bf q} \cdot (-{\bf k})}{ k \vert {\bf q} -  {\bf k} \vert}  \, ,
\end{eqnarray}
which leads us to 
\begin{eqnarray}
\label{c01}
&& \cos \beta_{{\bf k}, {\bf k} - {\bf q}} = \cos \beta_{-{\bf k}, -{\bf k} + {\bf q}}  \, .
\end{eqnarray}
 Condition \eqref{c01} clearly holds true for $\cos^2 \beta_{{\bf k},{\bf k}'}$ and for $\sin^2 \beta_{{\bf k},{\bf k}'}$. The substitution  $k \rightarrow -k$, equivalent to $\theta \longrightarrow \pi + \theta$, could always be used since we perform  an angular integration $\int\limits_0^{2 \pi} d \phi$.  To summarize, we write

\begin{equation}
\mbb{O}_{\gamma_1,\gamma_2}(\beta_{{\bf k},{\bf k}'}) \xrightarrow[\gamma_1 \leftrightarrow \gamma_2]{{\bf k} \leftrightarrow 
{\bf k}'} \mbb{O}_{\gamma_1,\gamma_2}(\beta_{{\bf k},{\bf k}'}) \, , 
\end{equation}
i.e., there is a complete reciprocity in any transitions $\gamma_1 \leftrightarrow \gamma_2$ and for ${\bf k} \leftrightarrow {\bf k}'$.

\medskip 
\par
The angular integration yields (in  units of $2 \pi/(\hbar v_F)$)

\begin{eqnarray}
&& \, \chi^{(+)}_{\phi}(q,T) = - \mu(T) -  k_B T \, \log\left\{
1 + \tet{exp} \left[
\frac{-\mu}{k_B T}
\right]
\right\} + \hbar v_F 
\int\limits_0^{q/2} \, dk \left\{
1+ \tet{exp} \left[
\frac{\hbar v_F k - \mu}{k_B T}
\right]
\right\}^{-1}   \\
\nonumber
&& \times \frac{1 +  2 \mc{B}_\phi -  (2 k / q)^2}{\sqrt{1 - (2 k / q)^2}} +
2 \hbar v_F \mc{B}_\phi \int\limits_0^{q/2} dk \, \left\{
1+\tet{exp}\left[
\frac{\hbar v_F k - \mu}{k_B T}
\right]
\right\}^{-1} \, [q/(2 k)]^2 \, \left\{ 1 - 
\left[1-(2k/q)^2\right]^{-1/2}
\right\}    \\
\nonumber 
&& + 2 \hbar v_F \mc{B}_\phi \int\limits_{q/2}^{\infty} dk \, 
\left(\frac{q}{2 k }\right)^2
\,
\left\{
1+\tet{exp}\left[
\frac{\hbar v_F k - \mu}{k_B T}
\right] 
\right\}^{-1}
 \,  ,
\end{eqnarray}

where $\mc{A}_\phi = 1/2 \, [1 + \cos^2 (2 \phi)]$ ranges between $\mc{A}_0 = 1$ for graphene and $\mc{A}_1 = 1/2$ for a dice lattice, and $\mc{B}_\phi = 1/2 \, \sin^2 (2 \phi) = 1 - \mc{A}_\phi$.  Also,

\begin{eqnarray}
&& \, \chi^{(-)}_{\phi}(q,T) = -(1 + 4 \mc{B}_\phi) \frac{\pi}{8} \, \hbar v_F q - k_B T \, \log\left\{
1 + \tet{exp} \left[
\frac{-\mu}{k_B T}
\right]
\right\} + \hbar v_F 
\int\limits_0^{q/2} \, dk \left\{
1+ \tet{exp} \left[
\frac{\hbar v_F k + \mu}{k_B T}
\right]
\right\}^{-1}     \\
\nonumber
&& \times \frac{1 +  2 \mc{B}_\phi -  (2 k / q)^2}{\sqrt{1 - (2 k / q)^2}} +
2 \hbar v_F \mc{B}_\phi \int\limits_0^{q/2} dk \, \left\{
1+\tet{exp}\left[
\frac{\hbar v_F k + \mu}{k_B T}
\right]
\right\}^{-1} \, [q/(2 k)]^2 \, \left\{ 1 - 
\left[1-(2k/q)^2\right]^{-1/2}
\right\}   \\
\nonumber 
&& + 2 \hbar v_F \mc{B}_\phi \int\limits_{q/2}^{\infty} dk \, 
\left(\frac{q}{2 k }\right)^2
\,
\left\{
1+\tet{exp}\left[
\frac{\hbar v_F k + \mu}{k_B T}
\right] 
\right\}^{-1}
 \,  ,
\end{eqnarray}
while 

\begin{equation}
\chi^{(0)}_{\phi}(q,T) = 0 \, . 
\end{equation}

Here, we used the following identities

\begin{equation}
\mbb{I}_{1}^{(\beta)}  = \int\limits_0^{2 \pi} \, \frac{\cos \phi \, d \phi}{c_1 + c_2 \,\cos \phi}  = \frac{1}{c_2}
\left[ 2 \pi - c_1 \, \mbb{I}_{0}^{(\beta)}
\right]\, ,\hskip0.1in c_2 > 0 \, , 
\end{equation}
as well as 
\begin{equation}
\mbb{I}_{2}^{(\beta)} = \int\limits_0^{2 \pi} \, \frac{\cos^2 \phi \, d \phi}{c_1 + c_2 \,\cos \phi} = \frac{c_1}{c_2^2}
\left[ - 2 \pi + c_1 \, \mbb{I}_{0}^{(\beta)}
\right]\, ,\hskip0.1in c_2 > 0 \, , 
\end{equation}
and for a general power $n$ 
\begin{equation}
\mbb{I}_{n}^{(\beta)} = \int\limits_0^{2 \pi} \, \frac{\cos^n \theta \, d \phi}{c_1 + c_2 \,\cos \phi} = 
\frac{(-1)^n}{c_2} \, 
\left(\frac{c_1}{c_2}\right)^{n-1} \,
\left[2 \pi - c_1 \, \mbb{I}_{0}^{(\beta)}
\right]\, ,\hskip0.1in c_2 > 0 \, , 
\end{equation}

expressed through a well-known integral

\begin{equation}
\mbb{I}_{0}^{(\beta)}  = \int\limits_0^{2 \pi} \, \frac{d \phi}{c_1 + c_2 \,\cos \phi} = \frac{2 \pi}{\sqrt{c_1^2 - c_2^2}} \,
\Theta \left(c_1^2 - c_2^2 \right) \,\sign(c_1 - c_2) \, , 
\end{equation}
which was widely used for similar calculations in graphene\,\,\cite{gonccalves2016introduction}. Here, we need to ensure that $\vert c_1 \vert > \vert c_2 \vert$ meaning that there are no poles on the real axis inside the unit circle. If this condition is not satisfied, the principal value of the integral is zero whereas it is actually divergent.  

\medskip 
\par
At zero temperature, 

\begin{eqnarray}
&& \, \chi^{(+)}_{\phi}(q,T \rightarrow 0 ) = - E_F + (1 + 4 \mc{B}_\phi) \frac{\pi}{8} \, \hbar v_F q - \mc{B}_\phi \frac{(\hbar v_F q)^2}{2 E_F} \,  ,
\end{eqnarray} 
for $q < 2 k_F$ and

\begin{eqnarray}
\chi^{(+)}_{\phi}(q,T \rightarrow 0) = && - E_F + \frac{E_F}{2} \sqrt{1 - \left( \frac{2 k_F}{q}\right)^2} + \hbar v_F q \, \left(\frac{1}{4} + \mc{B}_\phi
\right) \, \arcsin \left( 
\frac{2 k_F}{q} \right)   \\
\nonumber 
&& - \mc{B}_\phi \frac{(\hbar v_F q)^2}{2 E_F} \left[ 1 - \sqrt{1 - \left( \frac{2 k_F}{q}\right)^2} \, \right]
 \,  
\end{eqnarray}  
for $q > 2 k_F$. Thus, $\chi^{(+)}_{\phi}(q,T)$ could be presented as a single expression 

\begin{eqnarray}
\chi^{(+)}_{\phi}(q,T \rightarrow 0 ) = &&  - E_F + (1 + 4 \mc{B}_\phi) \frac{\pi}{8} \, \hbar v_F q - \mc{B}_\phi \frac{(\hbar v_F q)^2}{2 E_F}  +
\left\{ 
\frac{1}{2}\,\left(E_F + \mc{B}_\phi \, \frac{q^2}{E_F}  \right) \sqrt{1 - \left( \frac{2 k_F}{q}\right)^2}  \right. \\
\nonumber
&& - \left. \hbar v_F q \, \left[\frac{1}{4} + \mc{B}_\phi
\right] \, \arccos \left( 
\frac{2 k_F}{q} \right) \, 
\right\} \, \Theta(q - 2 k_F)
 \,  .
\end{eqnarray} 

The hole contribution term $\chi^{(-)}_{\phi}(q,T)$ at $T \rightarrow 0$ becomes

\begin{eqnarray}
&& \, \chi^{(-)}_{\phi}(q,T \rightarrow 0) = -(1 + 4 \mc{B}_\phi) \frac{\pi}{8} \, \hbar v_F q \,  ,
\end{eqnarray}
which does not have a special point (discontinuity of its derivative) at $q =  2 k_F$.

\color{black}
\clearpage
\bibliography{ATBib}

\end{document}